# A novel method and comparison of methods for constructing Markov bridges

F. Baltazar-Larios[1*] and Luz Judith R. Esparza[2†]

[1*]Facultad de Ciencias, Universidad Nacional Autónoma de México, México.
[2]Investigadora por México, Universidad Autónoma de Aguascalientes, México.

*Corresponding author(s). E-mail(s): fernandobaltazar@ciencias.unam.mx;
Contributing authors: judithr19@gmail.com;
[†]These authors contributed equally to this work.

**Abstract**

In this study, we address the central issue of statistical inference for Markov jump processes using discrete time observations. The primary problem at hand is to accurately estimate the infinitesimal generator of a Markov jump process, a critical task in various applications. To tackle this problem, we begin by reviewing established methods for generating sample paths from a Markov jump process conditioned to endpoints, known as Markov bridges. Additionally, we introduce a novel algorithm grounded in the concept of time-reversal, which serves as our main contribution. Our proposed method is then employed to estimate the infinitesimal generator of a Markov jump process. To achieve this, we use a combination of Markov Chain Monte Carlo techniques and the Monte Carlo Expectation-Maximization algorithm. The results obtained from our approach demonstrate its effectiveness in providing accurate parameter estimates. To assess the efficacy of our proposed method, we conduct a comprehensive comparative analysis with existing techniques (Bisection, Uniformization, Direct, Rejection, and Modified Rejection), taking into consideration both speed and accuracy. Notably, our method stands out as the fastest among the alternatives while maintaining high levels of precision.

# 1 Introduction

Markov jump processes (MJP) can be used to model systems in areas such as finance ([1]), epidemiology ([2], [3], [4]), population ecology ([5],[6]), biology ([7],[8]), and genetic ([9],[10],[11]). To model a system using MJP it is necessary to know its parameters, otherwise we have to make inference of them. For the case when a continuous record of such a MJP has been observed (ideal case), statistical inference is straightforward, since the sufficient statistics are simply the number of transitions between any two states and the total time spent in each state ([12], [13], [14]). However, if a MJP is only observed at discrete time points the inference problem become very complex. This situation corresponds to a missing data problem, for which the Expectation-Maximization (EM) algorithm and Markov Chain Monte Carlo (MCMC) methods are classical tools to make the inference ([1]).

For statistical inference in the case of incomplete observations, we have to calculate the sufficient statistics conditioned on discrete time observations. In this situation, it is important to have efficient methods for simulating paths of the MJP that start with the observed value and end at the time of the next value of a sample observation. The resulting conditioned sample is called Markov bridge (MB). MBs provide a way to condition a MJP to start and end at specific states ([15]). This is useful in applications where the initial and final conditions are critical, such as in modelling stock prices, where the process needs to start and end at specific prices. Moreover, MBs play a crucial role in estimating parameters of the underlying process. This is important in various scientific and engineering fields where understanding and estimating system dynamics are essential.

The applicability of MJPs in the context of discretely-sampled data has facilitated their integration into interdisciplinary literature, e.g., the reference [16] applied the problem of sampling path from an endpoint-conditioned process in the field of mathematical finance, while [17] applied to molecular evolution.

Various methods for sampling MBs are found in the literature, including direct sampling ([18]), uniformization ([19]), bisection ([9]), rejection sampling, modified rejection sampling ([20]), among others. Some approaches, albeit with limitations, involve an eigenvalue decomposition of the rate matrix ([21]), results for integrals of matrix exponentials ([22]), Laplace transforms, and continued fractions ([23]). A comparison of the last three methods can be found in [11]. Additionally, alternative methods for constructing MBs are proposed in [24], [25], and [26].

The construction of a MB hinges on two pivotal factors: precision and time efficiency. The effectiveness of each method in these dimensions is contingent on the structure of the intensity matrix (infinitesimal generator), the number of states, and the length of the MB, denoted as the distance between two observations. Given that, in numerous applications of MJP, the intervals between observations are often substantial, the importance of possessing a method that is both accurate and efficient in estimating MJP parameters becomes notably significant.

In this paper, we introduce a novel method to simulate a MB with these characteristics. The approach involves sampling a MB by simulating two processes independently: one moving forward in time from the initial point to the end time and another, time-reversed process moving backward in time from the end state to the initial time. If the sample paths of the processes intersect, we can combine them to obtain a new MB. We will call this: the time-reverse (TIR) method.

We present the theoretical foundation for the TIR method. To illustrate its accuracy and speed, we provide an example involving statistical inference for a discretely observed MJP using the TIR method. The estimators are obtained through the Monte Carlo Expectation-Maximization (MCEM) algorithm and the MCMC method. Furthermore, we conduct a comparison of both accuracy and speed among different methods, including direct sampling (DIR), the uniformization (UNI) method, the bisection (BIS) method, rejection sampling (REJ) method, modified rejection sampling (MOR) method, and the TIR method.

Hence, the principal contributions of this paper encompass:

*i)* Introducing a novel algorithm for simulating Markov Bridges.
*ii)* Demonstrating the accuracy and speed of the proposed algorithm.
*iii)* Establishing the ease of implementation of this algorithm.

The structure of this paper unfolds as follows. In Section 2, we elucidate the motivation behind the need for efficient methods in constructing MBs to facilitate statistical inference for discretely observed MJPs. Within this section, we introduce the MCEM algorithm and the Gibbs sampler for addressing these inference challenges. Additionally, we provide a succinct overview of the five most significant methods for generating MBs, delineating their respective advantages and disadvantages. In Section 3, we present the time-reverse algorithm for the generation of MBs. Section 3 entails a simulation study where the Time-Reversed (TIR) method is employed to simulate MBs, and subsequently, the infinitesimal generator of the MJP is estimated using the EM-algorithm and Gibbs sampler. Section 4 is dedicated to a comprehensive assessment of the accuracy and speed, conducting a comparative analysis of various methods, namely: DIR, UNI, BIS, REJ, MOR, and TIR. Finally, in Section 5, we present the concluding remarks.

## 2 Background and motivation

Consider a system evolution that can be represented as a time-homogeneous MJP, which is positive recurrent[1] and possesses a finite state space $E = \{1, 2, \ldots, n\}$ (an ergodic process). This model is characterized by the infinitesimal generator $\mathbf{\Lambda} = (\lambda_{ij})_{i,j \in E}$, with $\lambda_i := -\lambda_{ii}$, and the corresponding transition semigroup $\{P(t)\}_{t \geq 0}$, with elements $p_{ij}(t)$. These conditions ensure the existence of the stationary distribution $\boldsymbol{\pi} = (\pi_i)_{i \in E}$. For detailed definitions of these concepts, refer to [?].

**Remark 2.1.** *For all $i, j \in E$, we have that the stationary distribution $\boldsymbol{\pi}$ satisfies that*

$$p_{ij}\pi_i = p_{ji}\pi_j.$$

Let $X_t$ be the state of the system at time $t$, and $\boldsymbol{X} = \{X_t\}_{t=0}^{T}$ the values of the system between the times 0 and $T$. In order to model the evolution of the system, it is necessary to make inference of $\boldsymbol{\Lambda}$ based on the observations (states) of the system within the time interval $[0, T]$.

---

[1] A MJP is positive recurrent if all its states have a finite mean recurrence time.

## 2.1 Inference for MJP

The maximum likelihood estimation of the infinitesimal generator $\boldsymbol{\Lambda}$ from a complete observation of the process $\boldsymbol{X}$ is straightforward (see [14]). The likelihood function when $\boldsymbol{X}$ has been observed continuously and conditioned on the initial state is given by (see [12]):

$$L_T^c(\boldsymbol{\Lambda}) = \prod_{i=1}^n \prod_{j \neq i}^n \lambda_{ij}^{N_{ij}(T)} \exp(-\lambda_{ij} R_i(T)), \tag{1}$$

where $N_{ij}(T)$ is the number of jumps from state $i$ to state $j$, and $R_i(T)$ is the total time spent in the state $i$ in $[0, T]$. Thus, the sufficient statistics of the MJP are the time spent in each state $(R_i(T))$ and the number of jumps between any two states $(N_{ij}(T))$. It is not difficult to see that the maximum likelihood estimator (MLE) of the $ij$-th element of $\boldsymbol{\Lambda}$ is given by

$$\hat{\lambda}_{ij}(T) = \frac{N_{ij}(T)}{R_i(T)}, \tag{2}$$

provided that $R_i(T) > 0$ for all $i \in E$.

Suppose the MJP is discretely observed in a set of points $\{0 = t_0, t_1, \ldots, t_m = T\}$, for some $m \in \mathbb{N}$; and let $\boldsymbol{Y} = \{X_{t_0} = x_0, X_{t_1} = x_1, \ldots, X_{t_m} = x_m\}$ be the corresponding observations. In [27], the authors demonstrated that the MLE of $\boldsymbol{\Lambda}$ based on $\boldsymbol{Y}$ can be found either by the EM algorithm or by a MCMC procedure. In both cases, in order to find the estimators, we must have to calculate the sufficient statistics conditioned on the endpoints $X_{t_{i-1}} = x_{i-1}$ and $X_{t_i} = x_i$, for $i = 1, \ldots, m-1$. In this paper, to implement these methodologies, we need to simulate paths of the MJP conditioned on the endpoints; for this purpose we need the following definition.

**Definition 2.1.** *A Markov bridge with parameters $s, a, b$ (denoted by $(0, a, s, b)$-MB) is a stochastic process with time parameter $t \in [0, s]$, and with the same distribution as the MJP $\{X_t\}_{t=0}^s$ conditioned on $X_0 = a$ and $X_s = b$ with $a, b \in E$ and $s > 0$.*

Let $\Delta = t_i - t_{i-1}$ for all $i = 1, \ldots, m$. The MCEM-algorithm and the MCMC procedure are going to be briefly explained.

### 2.1.1 MCEM-algorithm

To optimize the likelihood function, the MCEM algorithm operates in the following manner.

**Algorithm 1** MCEM-algorithm for estimating $\mathbf{\Lambda}$.
────────────────────────────────────────────
1: Choose an initial value $\mathbf{\Lambda}^0$ and make $k = 0$.
2: Sample a continuous path $\mathbf{X} = \{X_t\}_{t=0}^T$, using the current $\mathbf{\Lambda}^k$ conditioned on $\mathbf{Y}$.
3: **E-step** Calculate the sample counterparts $\hat{N}_{ij}(T)$ and $\hat{R}_i i(T)$ of the sufficient statistics $N_{ij}(T)$ and $R_i(T)$ using $\mathbf{X}$ where

$$\hat{N}_{ij}(T) = \mathbb{E}[N_{ij}(T)|\mathbf{\Lambda}^k, \mathbf{Y}] = \sum_{i=1}^m \mathbb{E}[N_{ij}(\Delta)|\mathbf{\Lambda}^k, x_{i-1}, x_i],$$

and

$$\hat{R}_i(T) = \mathbb{E}[R_i(T)|\mathbf{\Lambda}^k, \mathbf{Y}] = \sum_{i=1}^m \mathbb{E}[R_i(\Delta)|\mathbf{\Lambda}^k, x_{i-1}, x_i].$$

4: **M-step** Calculate the estimators

$$\hat{\lambda}_{ij}^{k+1} = \frac{\hat{N}_{ij}(T)}{\hat{R}_i(T)}.$$

5: Update

$$\mathbf{\Lambda}^{k+1} = (\hat{\lambda}_{ij}^{k+1})_{i,j \in E}.$$

6: $k = k + 1$ and go to 2.
────────────────────────────────────────────

The steps 2-6 of the Algorithm 1 are repeated until convergence. In the E-step, we propose to use a MCMC method to estimate the conditional expectations. Then, to implement the E-step, we need to simulate paths of $\{X_t\}_{t=t_{i-1}}^{t_i}$ conditioned on $X_{t_{i-1}} = x_{i-1}$ and $X_{t_i} = x_i$, i.e., we have to simulate a $(0, x_{i-1}, \Delta, x_i)$-MB, for $i = 1, \ldots, m$.

In the work by Bladt et al. [28], an alternative method for calculating $\hat{N}_{ij}(T)$ and $\hat{R}_i(T)$ in the E-step is proposed. The authors demonstrated that these expressions satisfy some differential equations, and they numerically solved them using a fourth-order Runge-Kutta method.

### 2.1.2 Markov Chain Monte Carlo approach

For the MCMC method, we can apply the Gibbs sampler (see [29]) to the likelihood function (1). A natural prior is the gamma distribution given by

$$p(\mathbf{\Lambda}) \propto \prod_{i=1}^n \prod_{j \neq i}^n \lambda_{ij}^{a_{ij}-1} e^{-\lambda_{ij} b_i}, \tag{3}$$

where $a_{ij}, b_i > 0$ are constants to be chosen conveniently. The posterior distribution is given by

$$p^*(\boldsymbol{\Lambda}) = L_T^c(\boldsymbol{\Lambda})p(\boldsymbol{\Lambda}) \propto \prod_{i=1}^n\prod_{j\neq i}^n \lambda_{ij}^{N_{ij}(T)+a_{ij}-1} e^{-\lambda_{ij}(R_i(T)+b_i)}. \qquad (4)$$

The Gibbs sampler works as follows.

---
**Algorithm 2** Gibbs sampler for estimating $\boldsymbol{\Lambda}$.

---
1: Draw an initial $\boldsymbol{\Lambda}^0$ from the prior distribution (3) and make $k=0$.
2: Sample a continuous path of $\boldsymbol{X} = \{X_t\}_{t=0}^T$, using the current $\boldsymbol{\Lambda}^k$ conditioned on $\boldsymbol{Y}$.
3: Calculate the statistics $N_{ij}(T)$ and $R_i(T)$.
4: Draw $\boldsymbol{\Lambda}^{k+1}$ from the posterior distribution (4) and make $k=k+1$.
5: Go to 2.

---

After a burn-in period $K_0$, we can obtain $K - K_0$ samples from $\boldsymbol{\Lambda}$ and

$$\hat{\boldsymbol{\Lambda}} = \frac{\sum_{k=K_0+1}^K \boldsymbol{\Lambda}^k}{K - K_0}.$$

In order to implement the Algorithm 2 the main issue is how to generate sample paths of $\{X_t\}_{t=t_{i-1}}^{t_i}$ conditioned on $\boldsymbol{Y}$, i.e., a $(0, x_{i-1}, \Delta, x_i)$-MB, $i = 1, \ldots, m$.

In scenarios where we have observed the evolution of a system by a large number of independently endpoint-conditioned MJP, the goal is to generate sample paths for a large number of combinations of start and ending points by Markov bridges. Thus, having efficient methodologies to simulate Markov bridges for any distance between endpoints becomes crucial and pertinent.

## 2.2 Methods for sampling Markov bridges

At least seven methods exist for constructing Markovian bridges. In [19] and [11], the authors have implemented three methods and compared them: Eigenvalue decomposition of the rate matrix (EVD), UNI, and integrals of matrix exponentials (EXPM). The results show that EVD is less general than the other two algorithms since it requires a diagonalizable rate matrix, and depends on the reversibility of the matrix; the three methods have similar accuracy and the EXPM method is the most accurate one; also the EVD and UNI are faster than EXPM.

In [10], the authors provided a detailed summary of some path sampling algorithms from an endpoint-conditioned MJP: REJ, DIR, UNI, and BIS; with a brief discussion of Metropolis-Hastings proposals of endpoint-conditioned histories. Based on [10], we will consider these four algorithms in order to make a comparison with the method that we are going to propose in this paper.

Now, we present a brief description of these methods including their advantages and disadvantages to generate samples of a $(0, a, T, b)$-MB (see Figure 1). We recommend the reader [10] for more details of these methods; and references [17, 30] to see comparisons of them.

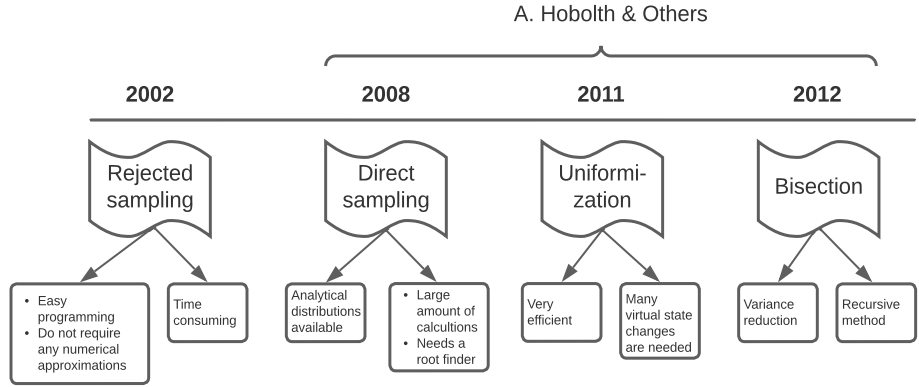

**Fig. 1** Main methods for simulating Markov bridges.

### 2.2.1 Rejection Sampling (REJ)

The REJ method is the least complex method of those presented here. Specifically, this algorithm simulates a trajectory of the MJP by conditioning the initial position $X_0 = a$, and finishes until generating the path that coincides with the final position $X_T = b$, rejecting all trajectories that did not meet this condition (see [16]).

In the REJ method the probability of hitting the observed ending state $b$ is $p_{ab}(T) := \mathbb{P}(X_T = b | X_0 = a)$. If $T$ is large, then $p_{ab}(T) \approx \pi_b$, and if $T$ is small then $p_{ab}(T) \approx \lambda_{ab}T$. Therefore, this method requires excessive execution time since many sample paths are rejected if:

- $T$ is large and $\pi_b$ is small, or,
- $T$ is small and $a \neq b$.

Therefore, the efficiency of rejection sampling increases as the space of valid endpoint conditions is enlarged [17].

### 2.2.2 Modified Rejection Sampling (MOR)

In order to minimize the execution time caused by the rejection of the samples generated by the REJ method, Rasmus Nielsen ([20]) developed the MOR method focusing
7

particularly on the case when the final state is different from the initial one ($a \neq b$) and there is a considerably short time horizon $T$.

Here, we have to simulate the time $\tau$ of the first state change ($c \neq a$) conditioned on at least one state change occurs before $T$ and $X_0 = a$. We construct a new $(0, c, T - \tau, b)$-MB using the REJ method if $b = c$ and the MOR method if $b \neq c$.

This modified sampling method explicitly avoids simulating constant sample paths when it is known that at least one state change take place. This method is very efficient if $T$ is small since the probability that there are no jumps (approximately $1 - \lambda_a T$) is large. On the other hand, if the transition probability from $a$ to $b$, given by $\frac{\lambda_{ab}}{\lambda_a}$, is small, this method is inefficient.

### 2.2.3 Direct method (DIR)

The DIR method was published in 2008 by Asger Hobolth ([18]). The idea of this method is to generate the stay time from the density function in terms of the spectral decomposition of the infinitesimal generator, as long as you change the state in the study interval. The most important condition to be met before performing this algorithm is that the infinitesimal generator accepts a decomposition by eigenvalues, i.e., we can write $\boldsymbol{\Lambda} = \boldsymbol{U}\boldsymbol{D}_\lambda \boldsymbol{U}^{-1}$, where $\boldsymbol{U}$ is the orthogonal matrix of eigenvectors, and $\boldsymbol{D}_\lambda$ is the diagonal matrix with the corresponding eigenvalues.

It is important to pay attention to the execution times of the DIR method, since this method is the most expensive computationally speaking due to the amount of calculations implemented to obtain the eigenvalues, and the numeric root finder implemented for finding the stay time in the visited states.

### 2.2.4 Uniformization (UNI)

The UNI method was first introduced by Arne Jensen and published in *Markov chains as an aid in the study of Markov processes* in 1953 ([31]).

For example, in 2006, [32] uses this algorithm to model the occurrence of a rare DNA motif. In 2011, Asger Hobolth and Jens Ledet Jensen ([19]) incorporated the Uniformization to simulate MB.

It is called Uniformization since the rate of moving from one state to another is the same for each of the states, and is calculated as $\mu = \max_i \lambda_i$. Thus, the algorithm forces the times of jumps following an exponential distribution with a uniform rate $\mu$ or, equivalently, the number of jumps in an interval of length $T$ follows a Poisson distribution with rate $\mu T$ causing the jump times to be evenly distributed over a time interval. The disadvantage of this method is the existence of virtual jumps, that is, jumps where there is no change of state, which occur with a rate $\mu - \lambda_k$ for some state $k$. The changes between states follow a discrete Markov chain with a transition matrix $\boldsymbol{\Gamma} = \boldsymbol{I} - \frac{1}{\mu}\boldsymbol{\Lambda}$, where $\boldsymbol{I}$ is the identity matrix.

### 2.2.5 Bisection (BIS)

The BIS method was implemented by Søren Asmussen and Asger Hobolth in [9]. Indeed, in 2008 the authors presented *Bisection ideas in End-Point Conditioned Markov Process Simulation*[2] where they showed the method.

The general idea of this method is to generate a recursive process where the simulation bridge ends with 0 or 1 jump:

1. If $X_0 = X_T = i$ and there no exists jumps, then $X_t = i$, for $0 \leq t \leq T$.
2. If $X_0 = i$ and $X_T = j$ and $i \neq j$ and there exist exactly one jump, then $X_t = i$ for $0 \leq t < \tau$ and $X_t = j$ for $\tau \leq t \leq T$.

The distribution of $\tau$ is given in the Lemma 4.1 and Remark 4.2 of the reference [9].

Since BIS is a recursive method, it becomes computationally expensive unlike its adversaries REJ, UNI, and MOR methods, which have a similar joint distribution. It is slightly more expensive to generate bridges over long distances than short ones, which is understandable since a greater distance implies a greater number of iterations.

## 3 Time-Reverse algorithm for Markov bridges

In this section, following the ideas of [33], we present a new algorithm for generating sample of MB that turns out to be very efficient when the distance between the observations is large.

Let $\boldsymbol{X^1} = \{X_t^1\}_{t=0}^T$ and $\boldsymbol{X^2} = \{X_t^2\}_{t=0}^T$ be two independent ergodic MJP with finite-state space $E = \{1, 2, \ldots, n\}$, infinitesimal generator $\boldsymbol{\Lambda}$, transition semigroup $\{P(t)\}_{t \geq 0}$, and stationary distribution $\boldsymbol{\pi}$, such that $X_0^1 = a$ and $X_0^2 = b$. The idea is to propose a novel method to sample a $(0, a, T, b)$-MB by simulating, in turn, the process $\boldsymbol{X^1}$ from $a$ forward in time, starting at time *zero* until time $T$, and the time-reversed process $\boldsymbol{\tilde{X}^2} := \{X_{T-t}^2\}_{t=0}^T$ of $\boldsymbol{X^2}$. If the samples paths of the processes intersect, we can combine these paths into a realization of the process to obtain a $(0, a, T, b)$-MB. Figure 2 presents a simulation example of $\boldsymbol{X}^1$ and $\boldsymbol{\tilde{X}}^2$ to obtain a MB.

Let consider the following results.

**Theorem 3.1.** *Let $\boldsymbol{X^1}$ and $\boldsymbol{X^2}$ be two independent MJP defined as before. Define $\tau = \inf\{0 \leq t \leq T : X_t^1 = \tilde{X}_t^2\}$ such that $\inf\{\emptyset\} = \infty$, and $\tilde{X}_t^2 = X_{T-t}^2$. Define the process $\boldsymbol{Z} = \{Z_t\}_{t=0}^T$ as follows:*

$$Z_t = \begin{cases} X_t^1, & if \quad 0 \leq t \leq \tau, \\ \tilde{X}_t^2, & if \quad \tau < t \leq T. \end{cases} \tag{5}$$

*Then, conditional on the event $\{\tau \leq T\}$, the distribution of $\boldsymbol{Z}$ equals the distribution of a $(0, a, T, b)$-MB, conditional on the event that the MB is hit by an independent MJP with state space $E$, infinitesimal generator $\boldsymbol{\Lambda}$, and initial distribution $p_{bk}(T)$ for all $k \in E$.*

---

[2]http://home.imf.au.dk/asmus

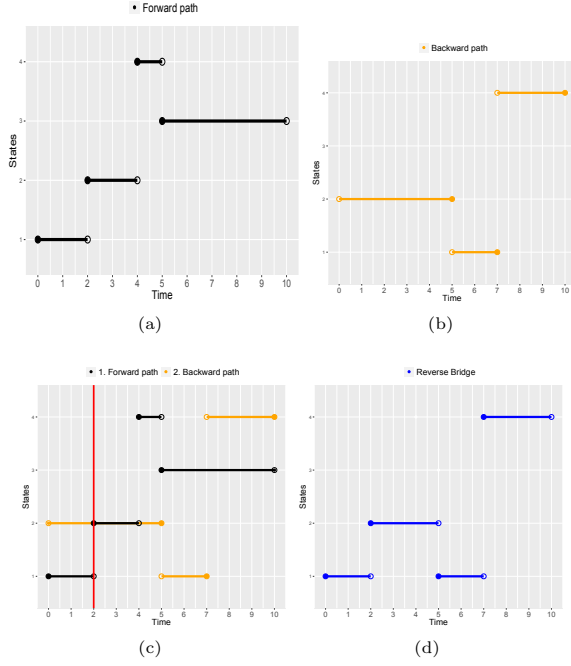

**Fig. 2** Considering $a = 1, b = 4$ and $T = 10$, the graphic (a) shows a $X_t^1$ path (black), the graphic (b) shows a $\tilde{X}_t^2$ path (yellow), in the graphic (c), we can see the intersection time between both paths given by $\tau = 2$ (red line). The graphic (d) shows the propose $(0, a, T, b)$-MB (blue).

Before proving the proof of the Theorem 3.1, we need to prove a lemma on the distribution of a time-reversed process.

**Lemma 3.1.** *The time-reversed process $\tilde{\boldsymbol{X}}^2 = \{\tilde{X}_t^2\}_{t=0}^T$ (defined as Theorem 3.1) has the same distribution as a MJP $\boldsymbol{X} = \{X_t\}_{t=0}^T$ with initial distribution $\boldsymbol{\pi}$ and $X_T = b$.*

*Proof.* For all $0 < s < t < T$ and $i, j \in E$, we have that the transition probability of $\tilde{\boldsymbol{X}}^2$ is

$$
\begin{aligned}
\tilde{p}_{ij}(t - s) := \mathbb{P}(\tilde{X}_t^2 = j | \tilde{X}_s^2 = i) &= \frac{\mathbb{P}(\tilde{X}_t^2 = j, \tilde{X}_s^2 = i)}{\mathbb{P}(\tilde{X}_s^2 = i)} \\
&= \frac{\mathbb{P}(X_{T-t}^2 = j, X_{T-s}^2 = i)}{\mathbb{P}(X_{T-s}^2 = i)} \\
&= \frac{\mathbb{P}(X_{T-s}^2 = i | X_{T-t}^2 = j)\mathbb{P}(X_{T-t}^2 = j)}{\mathbb{P}(X_{T-s}^2 = i)} \\
&= \frac{p_{ji}(t - s)\mathbb{P}(X_{T-t}^2 = j | X_0^2 = b)}{\mathbb{P}(X_{T-s}^2 = i | X_0^2 = b)}
\end{aligned}
$$

$$= \frac{p_{ji}(t-s)p_{bj}(T-t)}{p_{bi}(T-s)}$$

$$= \frac{\frac{\pi_i}{\pi_j}p_{ij}(t-s)\frac{\pi_j}{\pi_b}p_{jb}(T-t)}{\frac{\pi_i}{\pi_b}p_{ib}(T-s)} \quad \text{by Remark 2.1,}$$

$$= \frac{p_{ij}(t-s)p_{jb}(T-t)}{p_{ib}(T-s)}.$$

On the other hand, if $X_0 = k$ with $k \in E$, then

$$p_{ij}^{kb}(t-s) := \mathbb{P}(X_t = j | X_0 = k, X_s = i, X_T = b)$$

$$= \frac{\mathbb{P}(X_t = j | X_s = i)\mathbb{P}(X_T = b | X_t = j)}{\mathbb{P}(X_T = b | X_s = i)}$$

$$= \frac{p_{ij}(t-s)p_{jb}(T-t)}{p_{ib}(T-s)} = \tilde{p}_{ij}(t-s).$$

Hence $\tilde{\boldsymbol{X}}^2$ and $\boldsymbol{X}$ have the same transition probabilities.

Since the initial distribution of $\boldsymbol{X}$ is given by the stationary distribution $\boldsymbol{\pi}$, then so is for $X_T$, i.e., $\mathbb{P}(X_T = b) = \pi_b$. Therefore the conditional probability of $X_0$ given $X_T = b$ is

$$\mathbb{P}(X_0 = k | X_T = b) = \frac{\mathbb{P}(X_0 = k, X_T = b)}{\mathbb{P}(X_T = b)} = \frac{p_{kb}(T)\pi_k}{\pi_b} = \frac{p_{bk}(T)\pi_b}{\pi_b} = p_{bk}(T),$$

and the initial distribution of $\tilde{\boldsymbol{X}}^2$ is given by

$$\mathbb{P}(\tilde{X}_0^2 = k) = \mathbb{P}(X_T^2 = k) = \mathbb{P}(X_T^2 = k | X_0^2 = b)\mathbb{P}(X_0^2 = b) = p_{bk}(T),$$

then, $\tilde{\boldsymbol{X}}^2$ and $\boldsymbol{X}$ have the same initial distribution and we can conclude that they have the same distribution. $\square$

Now, we present the demonstration of our main result.

*Proof.* **of Theorem 3.1**. Let $\boldsymbol{X}^3$ be an ergodic MJP independent of the process $\boldsymbol{X}^1$, with infinitesimal generator $\boldsymbol{\Lambda}$ and initial distribution $\boldsymbol{\pi}$. We consider the Markov time $\kappa$ defined by

$$\kappa := \inf\{0 \leq t : X_t^1 = X_t^3\} \qquad \text{such that} \quad \inf\{\emptyset\} = \infty,$$

as the first time that the process $\boldsymbol{X}^3$ hits the process $\boldsymbol{X}^1$. If $\kappa \leq T$, we define the process $\boldsymbol{Y}$ by

$$Y_t := \begin{cases} X_t^1 & \text{if} \quad 0 \leq t \leq \kappa, \\ X_t^3 & \text{if} \quad \kappa < t \leq T. \end{cases}$$

Note that the processes $\boldsymbol{Y}$ and $\boldsymbol{X^1}$ have the same distribution (strong Markov property). Now, by Lemma 3.1, the processes $\boldsymbol{\tilde{X}^2}$ and $\boldsymbol{X^3}$ have the same distribution conditioned on $X_T^3 = b$, and then

$$\mathbb{P}(\boldsymbol{Y}|X_T^3 = b, \kappa \leq T) = \mathbb{P}(\boldsymbol{Z}|X_T^3 = b, \tau \leq T).$$

Since the event $\{Y_T = b, \kappa \leq T\}$ is a $(0, a, T, b)$-MB, so is $\boldsymbol{Z}$. □

Based on Theorem 3.1, we propose the following algorithm to draw paths of a $(0, a, T, b)$-MB.

---

**Algorithm 3** Reverse Method: Simulation of a $(0, a, T, b)$-Markov bridge

1: Simulate a MJP, $\boldsymbol{X^1}$, with intensity matrix $\boldsymbol{\Lambda}$ up to time $T$ such that $X_0^1 = a$. If $X_T^1 = b$, then $\boldsymbol{X^1}$ is a $(0, a, T, b)$-MB and stop. Otherwise go to Step 2.
2: Simulate a MJP, $\boldsymbol{X^2}$ independent of $\boldsymbol{X^1}$, with intensity matrix $\boldsymbol{\Lambda}$ up to time $T$ such that $X_0^2 = b$. Calculate $\boldsymbol{\tilde{X}^2}$, if $\tilde{X}_0^2 = a$ then $\boldsymbol{\tilde{X}^2}$ is a $(0, a, T, b)$-MB and stop. In other case, go to Step 3.
3: If there exists $\tau = \inf\{0 \leq t : X_t^1 = \tilde{X}_t^2\}$ such as $\tau \leq T$, then $\boldsymbol{Z}$ defined in (5), is a $(0, a, T, b)$-MB and stop. In other case, go to Step 1.

---

Algorithm 3 efficiently and rapidly simulates MBs, even in the context of extended time intervals, as elaborated in Section 4.

In the next section, a calibration study is carried out to verify that the proposed Algorithms 1 and 2 are yielding maximum likelihood estimators of the infinitesimal generator $\boldsymbol{\Lambda}$ in a satisfactory manner. We used simulated data and considered Algorithm 3 to sample Markov bridges.

sectionSimulation study

To illustrate the method for constructing Markov bridges proposed in this work, we examine an irreducible MJP $\{X_t\}_{t=0}^T$ with a state space $E = \{1, \ldots, 4\}$, but considering that direct communication between certain states is not possible, i.e., $\lambda_{i,j} = 0$ for some $i, j \in E$.

Specifically, in [11], the following model was considered, wherein the corresponding infinitesimal generator is non-reversible and possesses a complex decomposition:

$$\boldsymbol{\Lambda} = \begin{pmatrix} -4 & 2 & 1 & 1 \\ 0 & -3 & 2 & 1 \\ 1 & 0 & -3 & 2 \\ 2 & 1 & 1 & -4 \end{pmatrix},$$

where the initial distribution follows a uniform distribution.

We simulate a path of this MJP in the interval time $[0, 10]$. Assuming that we can observe the path at the discrete time points $0 = t_0 < t_1 < \ldots < t_{100} = 10$ with $\Delta = t_i - t_{i-1} = 0.1$, $i = 1, \ldots, 100$, then, we get the data $\{x_0, x_1, \ldots, x_{100}\}$.

We ran Algorithm 1 considering 150 iterations, and the burn-in occurred for all parameters in less than 50 iterations. To estimate the conditioned expectations in E-step, we considered 100 samples between each endpoints for each iteration. The initial values were $\lambda_{ij}^0 = 0.5$ for all $i,j$. We use Algorithm 3 to make the Markov bridges in the E-step of Algorithm 1.

For the Algorithm 2 the parameters for the prior distribution were $a_{ij} = b_i = 1$ for all $i,j$; we consider 500 iterations and a burn-in of 300 iterations. In this case, Algorithm 3 was used to generate the continuous path in step 2 of Algorithm 2.

The estimations using the MCEM and the MCMC are presented in Table 1. The estimator for MCEM is the average of the last 100 iterations and for MCMC is the average of the last 300 iterations. We also present 95% confidence intervals, i.e., the 0.025 and 0.975 quantiles of the corresponding sample.

**Table 1** Estimation using the MCEM and the MCMC algorithm including their confidence intervals (inferior CII and superior (CIS)).

| Parameter | True | MCEM estimation | | | MCMC estimation | | |
|---|---|---|---|---|---|---|---|
| | | Estimated | CII | CIS | Estimated | CII | CIS |
| $\lambda_{21}$ | 0 | 0.006 | 0.005 | 0.007 | 0.011 | -0.002 | 0.025 |
| $\lambda_{31}$ | 1 | 1.019 | 0.962 | 1.075 | 1.021 | 0.904 | 1.140 |
| $\lambda_{41}$ | 2 | 1.983 | 1.944 | 2.023 | 2.002 | 1.792 | 2.211 |
| $\lambda_{12}$ | 2 | 1.964 | 1.855 | 2.074 | 1.929 | 1.751 | 2.108 |
| $\lambda_{32}$ | 0 | 0.012 | 0.010 | 0.013 | 0.009 | -0.002 | 0.022 |
| $\lambda_{42}$ | 1 | 0.961 | 0.909 | 1.014 | 0.963 | 0.870 | 1.057 |
| $\lambda_{13}$ | 1 | 0.970 | 0.929 | 1.012 | 0.977 | 0.819 | 1.134 |
| $\lambda_{23}$ | 2 | 2.018 | 1.980 | 2.056 | 1.972 | 1.801 | 2.143 |
| $\lambda_{43}$ | 1 | 0.992 | 0.959 | 1.025 | 0.952 | 0.813 | 1.092 |
| $\lambda_{14}$ | 1 | 1.004 | 0.967 | 1.041 | 0.993 | 0.835 | 1.151 |
| $\lambda_{24}$ | 1 | 1.049 | 0.980 | 1.119 | 1.012 | 0.887 | 1.137 |
| $\lambda_{34}$ | 2 | 2.049 | 1.972 | 2.126 | 2.082 | 1.929 | 2.234 |

Table 1 shows that the true parameters using both methodologies are contained into their confidence intervals, except to the parameters equal to zero, that, given the nature of the MCEM algorithm, their corresponding confidence intervals do not contain their parameters.

Hence, we have calibrated our algorithms to identify effective estimation for the intensity matrix. However, it is worth noting that this task has been previously acknowledged and implemented in existing literature, as discussed earlier, employing several methods. In the following section, we will highlight the advantages of using the proposed algorithm for generating MBs.

## 4 Comparison

In this section, we provide a comparison of six methods employed for generating paths of MBs, considering both their accuracy and speed. The methods under consideration are UNI, BIS, DIR, REJ, MOR, and TIR.

## 4.1 Testing accuracy

Based on the reference [34], the conditional expectations of the sufficient statistics are given by:

$$\mathbb{E}[N_{ij}(t)|X_0 = x, X_t = y] = \frac{\lambda_{ij} I_{xy}^{ij}}{p_{xy}(t)} \tag{6}$$

and

$$\mathbb{E}[R_i(t)|X_0 = x, X_t = y] = \frac{\lambda_{ii} I_{xy}^{ii}}{p_{xy}(t)}, \tag{7}$$

where

$$I_{xy}^{ij}(t) = \int_0^t p_{xi}(s) p_{jy}(t-s) \mathrm{d}s, \tag{8}$$

for all $i, j, x, y \in E$, $t \geq 0$.

We will employ equations (6) and (7) to assess the accuracy of the methods.

Following [11], we consider a $n$-dimensional generator $\mathbf{\Lambda}$ with the following form:

$$\mathbf{\Lambda} = \begin{pmatrix} -1 & \frac{1}{n-1} & \frac{1}{n-1} & \cdots & \frac{1}{n-1} \\ \frac{1}{n-1} & -1 & \frac{1}{n-1} & \cdots & \frac{1}{n-1} \\ \vdots & \vdots & \vdots & \cdots & \vdots \\ \frac{1}{n-1} & \frac{1}{n-1} & \frac{1}{n-1} & \cdots & -1 \end{pmatrix}. \tag{9}$$

As $\mathbf{\Lambda}$ possesses two distinct eigenvalues: 0 with multiplicity 1 and $-\frac{n}{n-1}$ with multiplicity $n-1$, there exists an analytic solution for (8), as detailed in [11]. Consequently, we can compare the outcomes from the six methods against the true values of (6) and (7) by varying the number of states ($n = 3, \ldots, 20$) and employing random endpoints. It is crucial to emphasize that the applicability of the TIR method is restricted to cases where the process is stationary. For a given $\epsilon > 0$, we define the estimator of a stationary time as $\rho := \inf\{t > 0 : ||\Pi - P(t)|| < \epsilon\}$, where $\Pi$ represents the $n \times n$ matrix with all rows equal to the stationary distribution $\boldsymbol{\pi}$.

In Figure 3, we present the estimation of stationary times for different values of $n$, which are then used as the times between endpoints for the accuracy test.

In Figures 4 and 5, where $n$ represents the dimension of $\mathbf{\Lambda}$ ranging from 3 to 20, we depict the norm-1 of the conditional expectations of the sufficient statistics and their respective estimations obtained through all six algorithms.

In general, as observed in Figures 4 and 5, the TIR algorithm performs similarly to the others. It is challenging to discern which algorithm is better based solely on having a lower norm in both statistics. Efficiency is a relative measure that heavily depends on the specification of the conditional stochastic process. Computational requirements for each algorithm are contingent on factors such as the intensity matrix, time interval, and endpoints.

## 4.2 Testing speed

We will employ two models to assess the speed of the algorithms considering different form of the intensity matrix.

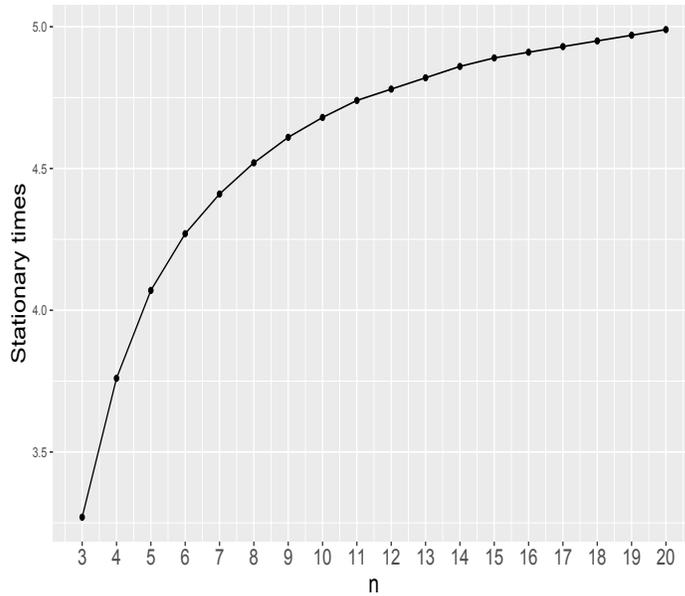

**Fig. 3** Estimation of the stationary times for $n = 3, \ldots, 20$.

The CPU execution time is calculated using a Mac Pro (Late 2013) with a 3.7GHz Quad-Core processor and 12 GB 1866 MHz DDR3 memory.

### 4.2.1 Model 1

Let us consider the MJP presented in Section 4.1, specifically when $n = 3$ and $T = 4, 5, \ldots, 9$. The stationary time for $n = 3$ is 3.25, as indicated in Figure 3; hence, we have selected values of $T$ greater than or equal to 4. Figure 6 illustrates the CPU execution times for the average of 1000 Markov bridges generated by each of the six methods with random endpoints.

Figure 6 illustrates that the Reverse, Modified Rejection, and Rejection methods consistently exhibit faster performance across all values of $T$, while the Direct method proves to be slower. The Bisection method falls between the speed of the Reverse and Direct methods. The Uniformization method demonstrates commendable execution times. Thus, having an intensity matrix in a manner that ensures an equal probability of transitioning to all states (9), our method yields results very similar to those already established in the existing literature. However, for practical purposes, modelling real phenomena that have such an intensity matrix is almost unlikely. Therefore, another example is presented considering another type of intensity matrix.

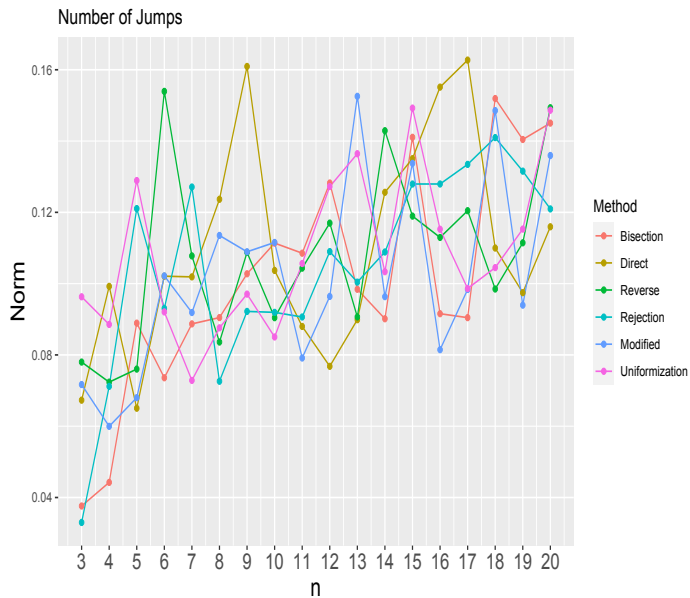

**Fig. 4** Accuracy test considering the number of jumps for the six methods.

### 4.2.2 Model 2

Let us extend the analysis of algorithm speed, but with a different form of the intensity matrix that involves more general interactions between states. In this scenario, we examine a MJP where each state has different parameters for stay times. Specifically, from state 1, transitions to other states occur with equal probability; from state 2, transition to state 3 occurs with probability one; and from state 3, transitions to other states happen with varying probabilities. Consider the MJP with an infinitesimal given by

$$\mathbf{\Lambda} = \begin{pmatrix} -2 & 1 & 1 \\ 0 & -10 & 10 \\ 4 & 1 & -5 \end{pmatrix}. \tag{10}$$

The stationary time of this MJP is 0.95. Consequently, in Figure 7, we present the CPU execution times for the average of 1000 Markov bridges using all six methods. This consideration involves random endpoints and spans across $T = 1, 2, \ldots, 6$.

As expected in this example, Figure 7 demonstrates that the TIR method is notably the fastest across all values of $T$. This is attributed to the stationary distribution $\boldsymbol{\pi} = (0.606, 0.091, 0.303)$. For bridges ending in state 2, where $\pi_2$ is the smallest, the methods REJ and MOR exhibit slower performance. Conversely, owing to $\mu - \lambda_1 = 10 - 2 = 8$, the UNI method involves numerous virtual jumps, resulting in a longer execution time.

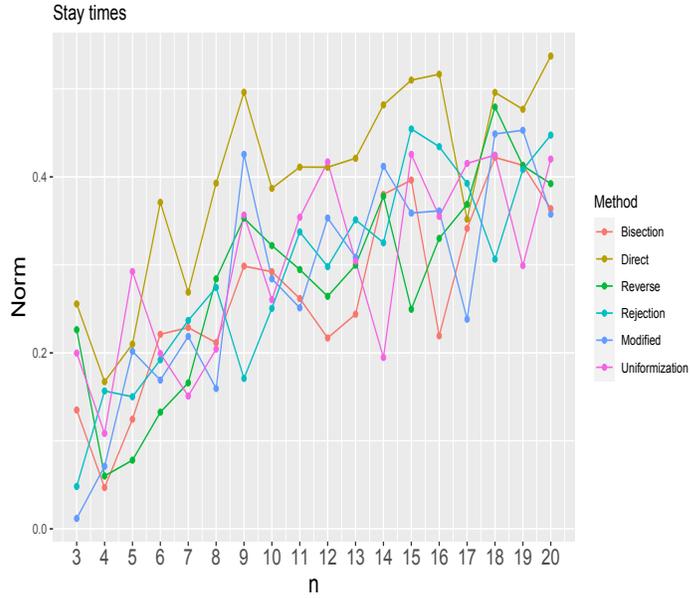

**Fig. 5** Accuracy test considering the stay times for the six methods.

Due to the virtual state changes inherent in the uniformization sampling process, the computational cost will generally be somewhat higher compared to other methods [17]. The choice between direct sampling and uniformization depends on the number of virtual state changes involved. In contrast, rejection sampling completes each iteration swiftly without relying on virtual transitions. The feasibility of rejection sampling is determined by the path acceptance probability. In the direct sampling algorithm, a root finder is utilized to simulate the waiting time before the next state change.

In general, it is most common for a system to be modeled by a MJP with an intensity matrix similar to (10) rather than (9). Following our analysis, it becomes evident that, given a sufficiently large distance between observations to ensure process stationarity, the TIR method stands out as the fastest while maintaining comparable precision to other existing methods.

## 5 Conclusions

The widespread use of MBs as a tool for inference in interdisciplinary studies has spurred the creation of some path-sampling algorithms. As the range of applications expands, there is an increasing demand for computationally efficient approaches.

MBs are versatile tools that find applications in diverse fields where modeling, simulation, and analysis of stochastic processes are essential for understanding system behavior, making predictions, and conducting statistical inference.

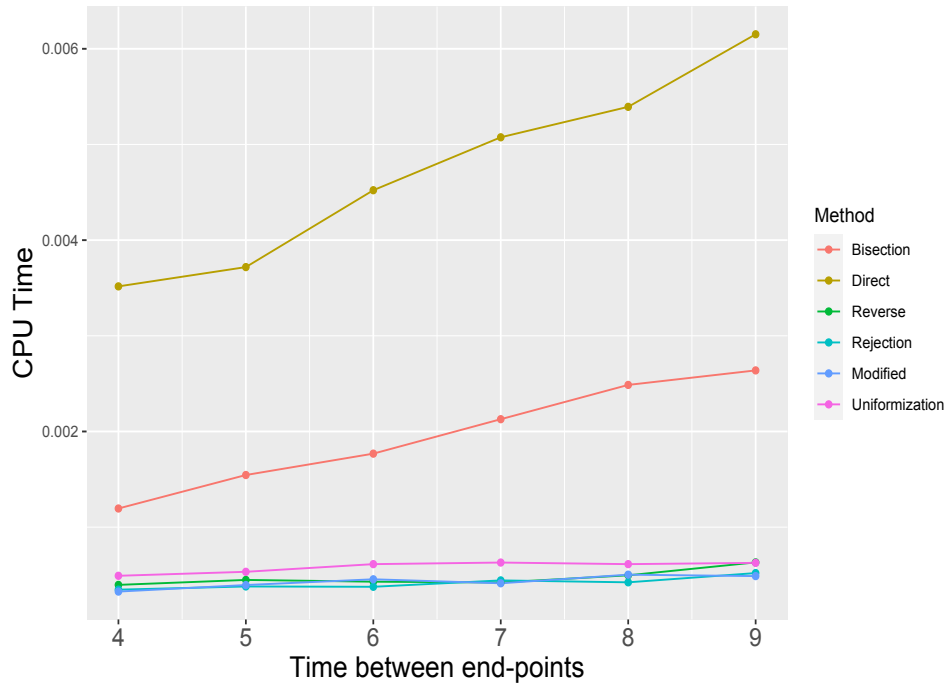

**Fig. 6** CPU time for sample paths generation using the Model 1 for $n = 3$ and $T = 4, 5, \ldots, 9$.

In mathematical modelling, having calibrated algorithms that can produce accurate estimates of model parameters is crucial. Over the years, several algorithms have been suggested for estimating parameters in models involving Markov jump processes, including direct sampling, uniformization, bisection, rejection sampling, and modified rejection sampling, among others. While all these methods have demonstrated their ability to provide reliable parameter estimates, the challenge lies in the execution time, which is heavily dependent on the intensity matrix's structure. Consequently, these algorithms might take a considerable amount of time to converge and obtain accurate estimators.

The computational demands of each algorithm hinge on factors such as the intensity matrix, the time interval $T$, and the endpoints. Generally speaking, no single algorithm outperforms the others, as each exhibits distinct strengths and weaknesses.

Hence, this article introduced a novel method for efficiently and rapidly simulating Markovian bridges. This approach ensures that the estimators of infinitesimal generators are just as accurate as those obtained through other methods.

However, the proposed method works particularly in scenarios where the distance between the starting and ending points is sufficiently large. To validate the relevance of this method, we conducted a simulation study on statistical inference for discretely observed Markov jump processes using both the MCEM algorithm and the MCMC method. Our findings provide evidence that, when applicable, this new method exhibits improvements over existing methods, considering key characteristics such as accuracy and running time.

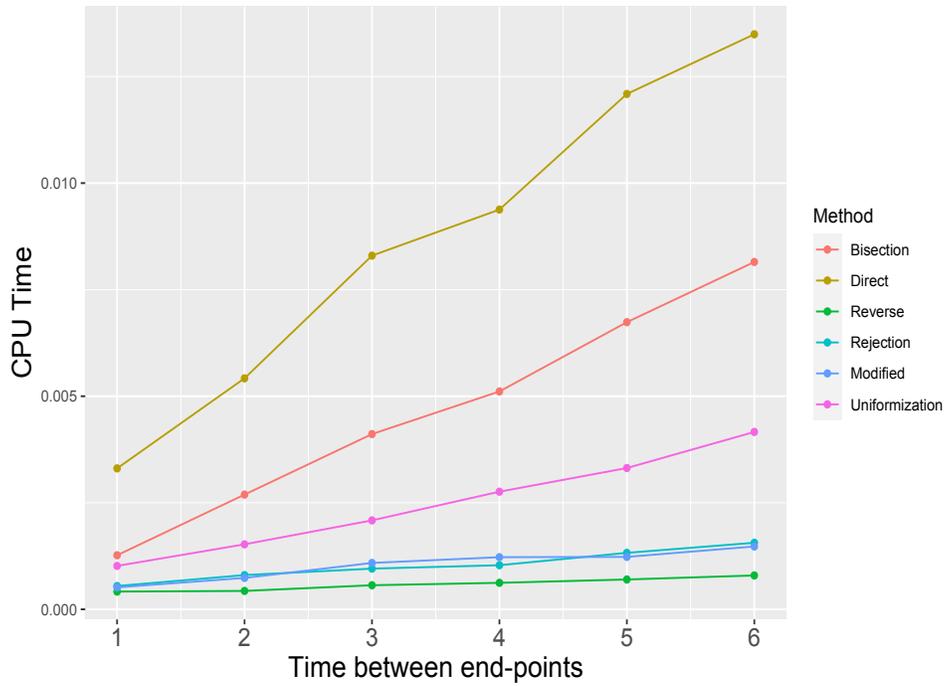

**Fig. 7** CPU time for sample paths generation using the Model 2 and $T = 1, 2, \ldots, 6$.

# Acknowledgments

We are very grateful to the two reviewers for many constructive comments and suggestions. LJRE is financially supported by the UAA grant PIM23-3.

# Appendix

## Abbreviations

| | |
|---|---|
| MJP | Markov Jump Processes |
| EM | Expectation-Maximization |
| MCMC | Markov Chain Monte Carlo |
| MB | Markov Bridge |
| TIR | Time Reverse |
| MCEM | Markov Chain Expectation-Maximization |
| DIR | Direct sampling |
| UNI | Uniformization |
| BIS | Bisection |
| REJ | Rejection sampling |
| MOR | Modified Rejection |
| MLE | Maximum Likelihood Estimator |
| EVD | Eigenvalue Decomposition |
| EXPM | Integrals of Matrix Exponentials |

## Program in R

https://github.com/judithr19/MarkovBridges